# Oxygen redox in battery cathodes: A brief overview

M. Hussein N. Assadi [1], Dorian A. H. Hanaor [2,3]

1. RIKEN Center for Emergent Matter Science (CEMS), 2-1 Hirosawa, Wako, Saitama 351-0198, Japan

2. Faculty of Science and Engineering, Southern Cross University, Military Road, East Lismore, NSW 2480, Australia

3. Institute of Materials Science and Technology, Technische Universität Berlin, 10623 Berlin, Germany

**Abstract:**

The participation of oxygen or other anionic species in redox activities in cathode materials for lithium and sodium-ion battery systems is known to play a role in governing the useful capacity of these batteries. Directly probing anionic redox mechanisms is not possible, rather the computational analysis by density functional theory poses the main approach towards gleaning insights into anionic redox activity and harnessing these effects to maximize capacity in future electrode materials. Here we showcase material systems exhibiting this mechanism of ion insertion and removal, and present the key computational considerations in studying anionic redox activities in battery materials. Aided by new computationally derived understandings of the role of anionic redox in emerging battery materials, increasingly greater levels of useable capacities can be extracted through informed materials design.

## 1  Introduction

The central feature of all rechargeable batteries is the reversible insertion and removal of an ionic species from electrode materials. Extracting higher performance levels from existing and future battery materials necessitates a deeper understanding and description of phenomena occurring in these materials during the insertion and removal of ions. Computationally derived insights also help us improve cyclability, enhance safety and shelf life, and replace expensive elements and synthesis methods with affordable and facile ones. Mainly, in cathode materials, the removal of a single $Li^+$ cation [1]—or any other cycled cations such as $Na^+$ [2] or $Mg^{2+}$ [3,4]—is most commonly accommodated chemically in the solid through the change in the oxidation state of a transition metal in a transition metal oxide (TMO). The most widely applied materials in these applications are metal oxides of the general form $LiTMO_2$, where TM represents a transition metal, commonly Ni, Mn, and Co.

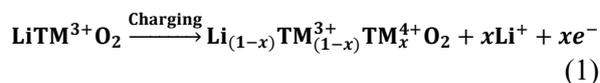

$$LiTM^{3+}O_2 \xrightarrow{\text{Charging}} Li_{(1-x)}TM^{3+}_{(1-x)}TM^{4+}_x O_2 + xLi^+ + xe^-$$
(1)

These metal oxides are primarily in the form of $\alpha$-$NaFeO_2$ type layered rocksalt structures [5]. Other common forms include layered hexagonal structures similar to those of $P2$-$NaCoO_2$ [6], $P3$-$NaMnO_2$ [7], and $O3$-$NaVO_2$ [8]. Depending on cationic species, decent theoretical capacities can potentially be achieved in these materials, typically in the range of 200~300 mAh/g, assuming that all lithium or sodium is de-intercalated during charging, which is rarely the case. In principle, this theoretical capacity is often unattainable because the de-lithiation process includes crystallographical phase changes that are not conducive to further alkali metal extraction [9]. Furthermore, the formation of the interphase materials, due to reactions with the electrolyte at the cathode surface, may create a diffusion barrier against the extraction of either Li or Na [10]. Cathode decomposition and oxygen gas release may also prevent battery systems from reaching their maximum theoretical capacity [11]. Despite all these limiting factors, the available capacities of actual cathode materials have achieved and even exceeded such levels of charge density, delivering an overcapacity exceeding the transition metal oxidation limit. The mechanisms through which this occurs at the atomic scale require further elucidation. In particular, the ability of oxygen, or other anions, to participate in redox reactions has been the subject of significant debate in the battery materials community since the early 2000s [12-14].

The concept of anionic redox, schematically shown in Fig. 1, was brought to the forefront of the field through studies regarding charge compensation mechanisms in $Li_2MnO_3$. During the charging of cathode materials based on $Li_2MnO_3$, lithium de-intercalates from the structure, which consists of alternating Li and [$Li_{1/3}Mn_{2/3}$] layers to the extent of 1.4 lithium ions per formula unit [1,5]. This occurs despite the occupied $t_{2g}$ orbitals, which represent the lower energy level of the split $d$-orbitals in +4 manganese cations, exhibiting a frontier energy level that is too low to undergo oxidation. Alternative mechanisms through which the requisite charge compensation may occur have been postulated, including the removal of oxygen or decomposition of the electrolyte solvent– which would impair reversibility and thus limit the cycling performance of such batteries. While an inevitable diminishment of capacity is often observed in cathodes after initial cycling due to irreversible or poorly reversible redox phenomena, reversible reactions through which oxygen species undergo redox reactions in the bulk of the cathode material offer pathways towards the reversible capacities observed in this and related transition metal oxides. This has emerged as a theme of interest in designing and interpreting high-density cathodes in Na- and Li-ion systems.

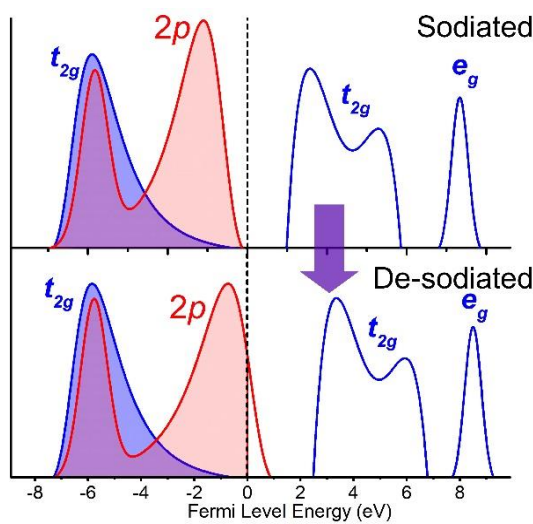

**Fig. 1.** The schematic representation of the O redox reaction decipherable from the electronic density of states. The upper panel shows a sodiated cathode TM oxide in which O redox is dominant, with O $2p$ states having a larger population near the Fermi level than those of the TM $d$ states. Upon desodation, electrons vacate $2p$ states, as they are more labile for electrochemical reactions.

## 2  Anionic Redox in Battery Electrode Materials

The redox activity of transition metal cations often sets the ultimate limit on the performance of electrode materials, primarily cathodes, in lithium- and sodium-ion batteries, accommodating the reversible insertion and removal of Li or Na. In both Li and Na ion batteries, transition metals' redox activity during discharging allows the reversible storage of electrons in the cathode. In many cathode materials, oxygen redox activity is known to take place. This involves oxide anions $O^{2-}$ donating electrons during charging, thus interestingly acting as reducing agents, which may compensate for the absence of oxidizable transition metals. This activity may allow greater energy storage capacity in battery electrodes than would otherwise be theoretically predicted and motivates its investigation through experimental and computational approaches.

The reversible redox activity of oxygen during charging can be summarised as the formation of peroxo-type oxygen species following

$$O^{2-} \longleftrightarrow (O_2)^{n-} \qquad (2)$$

Alternatively, invoking the formation of electron holes on oxygen atoms has been put forward to explain anion-driven excess capacity in alkali-rich cathode materials without the detrimental $O_2$ gas evolution. Localized configurations within a crystal structure will determine to what extent oxygen may partake in redox reactions. Labile oxygen states in Li/Na metal oxides arise from what has been termed Li-excess (or Na-excess) environments [15]. In traditional cathode materials with a nominal chemical composition of $LiTMO_2$ (or $NaTMO_2$), ions are arranged in TM∥O∥Li layers, as shown in Fig. 2a. TM ions are in octahedral coordination, implying that each TM ion has six nearest neighbor O ions. The same is true for Li (or Na) ions before de-intercalation. Moreover, O atoms are coordinated by three TM ions and three Li (Na) ions. Fig. 2b shows the hybridization between cations and oxygen in $LiTMO_2$-type materials. TM ions have five $d$ orbitals, one $s$ orbital, and three $p$ orbitals. As there are two O ions per any TM ion, there will be 6 O $2p$ orbitals to hybridize with the TM orbitals.

Under octahedral coordination, the M $d$ orbitals split into two groups, lower energy threefold $t_{2g}$, and higher energy twofold $e_g$ orbitals. The two $e_g$ orbitals along with the TM $s$ and TM $p$ orbital, hybridizing with the 6 O $2p$ orbitals, create 12 new molecular orbitals arranged in two sets; bonding and antibonding [16]. These new molecular orbitals are usually referred to as $p$-like orbitals. The bonding orbitals are two $e_g^b$, one $a_{1g}^b$, and three $t_{1u}^b$ orbitals. These orbitals are always occupied and located at the bottom of the valence band, *i.e.*, 8 ~ 10 eV below the Fermi level. The antibonding orbitals, denoted $e_g^*$, $a_{1g}^*$, and $t_{1u}^*$, mirror the bonding orbitals in character and are formed at about 10 to 12 eV higher. M's $t_{2g}$ orbitals, due to their orthogonal orientation to O's $2p$ orbitals, do not significantly hybridize with oxygen orbitals and are therefore located between the bonding and antibonding orbitals, closer to the Fermi level. Because of this lack of hybridization, the $t_{2g}$ orbitals are referred to as $d$-like orbitals in the solid. The interaction between Li (or Na) with oxygen is more ionic, leaving the alkali empty $s$ orbital way above the Fermi level, up to ~14 eV, and thus not directly participating in the redox chemistry.

Given the arrangement of the hybridized orbitals (Fig. 2b), the location of the Fermi level depends on the electronic population of the $TM^{2+}$ ion. If $TM^{2+}$ does not have any $e_g$ electron, for instance, low spin $Mn^{2+}$ or $Co^{2+}$, the Fermi level is located above the highest occupied $t_{2g}$ orbital. Consequently, during the charging process, electrochemical extraction of any electrons comes from the occupied $t_{2g}$ orbitals, as they are the closest to the Fermi level. In other words, they are electrochemically labile. This is why in most $LiTMO_2$ (or $NaTMO_2$) type cathode materials, the redox reaction is assumed to be borne on the TM ions. A caveat is that magnetic exchange usually removes the degeneracy between spin-up and spin-down $t_{2g}$ and $e_g$ orbitals. However, this exchange is not strong enough to alter the labile electrons for the redox reaction, but it influences the operating potential nonetheless [17].

For oxygen redox to occur, the orbitals with $p$ characters should be lifted closer to the Fermi level so they, too, become labile for electrochemical electron extraction and insertion. Such alteration to the orbital arrangement can be achieved by changing oxygen's coordination environment. For instance, as a major class of transition metal oxides used in Li and Na-ion battery electrodes show layered rocksalt-type structures, which may convert to spinel-type structures during cycling. In these layered rocksalt structures, the layer containing transition metal cations is often substituted with Li or Na ions. These materials are referred to as Li-excess (or Na-excess) materials. An essential feature in these materials, as shown in Fig. 2c, is that the oxygen ions, on average, are now coordinated by less than three TM ions. That is because each oxygen is still coordinated by three Li (or Na) ions from one side. However, from the other side, the layer now contains both alkali and transition metals. As a result, some oxygen atoms will also be coordinated by additional Li or Na ions from this layer.

As discussed earlier in conventional cathode materials with exclusive TM–O–Li coordination, the bonding hybridization with TM $s$ and $p$ orbitals pulls the bonding—and

occupied—*p*-like orbitals down to the bottom of the valence band. However, in Li-excess (or Na-excess) materials, there is no such hybridization for those oxygen ions with Li–O–Li (or Na–O–Na) coordination. As a result, the unhybridized or *orphaned* O 2*p* states are similar to those non-bonding $t_{2g}$ states in their energy level, both being less hybridized. A schematic of this orbital arrangement is given in Fig. 2d. Now that O 2*p* states are close to TM $t_{2g}$ states, both O and TM can potentially donate or accept electrons during the redox reaction. The exact contribution of O *vs.* TM to the redox reaction depends on many factors, such as the M–O bond's covalency, the TM ions' electronic population, and the delithiation (or desodiation) level [15].

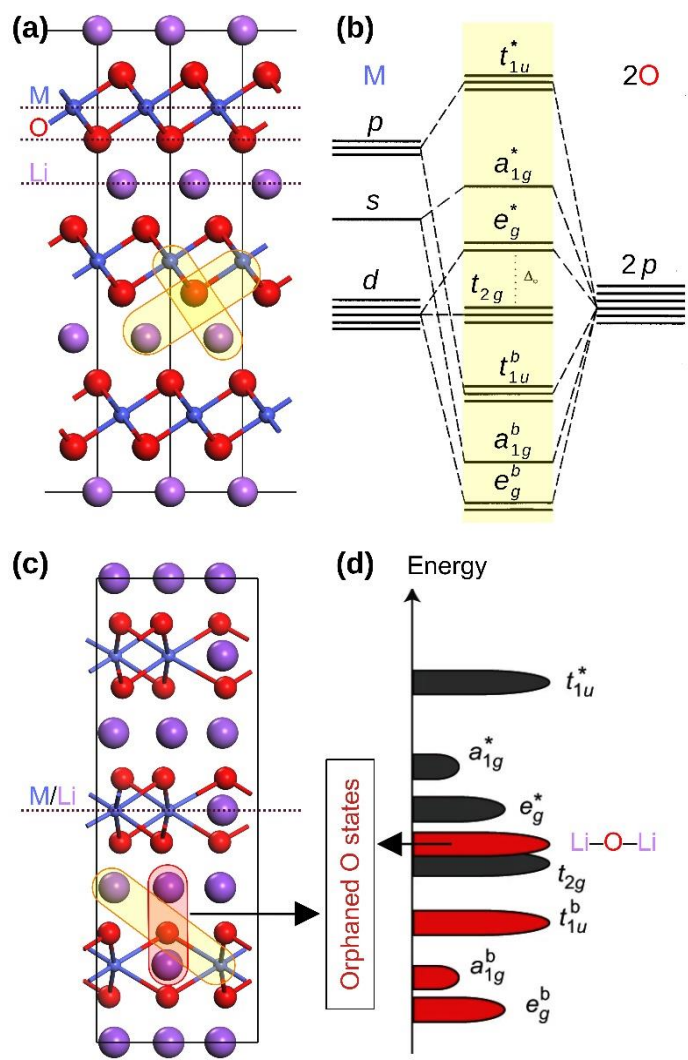

**Fig. 2.** (a) The structure of layered hexagonal *P*3-LiTMO$_2$, a common cathode material. Here, all oxygen ions are coordinated by three transition metal (TM) ions and three Li ions, for which examples are highlighted in yellow. (b) A schematic of molecular orbitals, highlighted in yellow, arising from the TM–O hybridization. (c) The structure of Li-excess layered hexagonal Li$_2$TMO$_3$. Here, the layer containing the TM ions also contains Li ions in random distribution. Consequently, some of the O ions are coordinated by more than three Li ions, for which an example is highlighted in red. (d) The molecular orbital arrangement in Li-excess Li$_2$TMO$_3$ and the position of the orphaned O states. (b) and (d) adapted from Refs [15,16]. Copyrights 1997 American Physical Society and 2016 Springer-Nature.

Certain electrode materials are known to be dominated by oxygen redox effects, for example, Li$_2$MnO$_3$, in which Manganese is known to maintain its +4 oxidation state and is generally not applied as a stand-alone cathode material in commercial batteries. As the transition metals sites are not observed in this cathode material to exhibit extensive redox activity, the performance of this material is attributed to oxygen redox activity [2,13,18,19]. Moreover, LiMn oxides are mostly modified in cation substitution to form the basis of $LiMn_xNi_{1-x}O_2$ (LMNOs), which form the backbone of the cathodes in many widely used battery systems. Cobalt is often included too, and indeed it is oxides with compositions including $Li_{1+x}Ni_yCo_zMn_{(1-x-y-z)}O_2$ where many of the first explorations of anionic redox effects have been studied [20]. In such systems, $Li_{1.2}Mn_{0.54}Co_{0.13}Ni_{0.13}O_2$ represents a commonly studied stoichiometry [21]. Its modification by various forms of cation substitution is pursued experimentally and through *ab initio* studies with a view towards suppressing layered-to-spinel type transformations and modulating anionic redox behavior.

## 3 Experimental and Theoretical Investigations

The identification of oxygen redox effects as playing a significant role in facilitating lithium de-intercalation in various $LiTMO_2$ systems inspires the search through experimental and computational methods for new cathode materials. Experimentally ascertaining the occurrence of oxygen redox effects in battery materials is not at all trivial. In general, both in sodium-ion and lithium-ion battery materials, the occurrence of oxide ion redox should increase the measurable capacity beyond what is possible through transition metal redox effects. However, proportionating battery capacity to effects observed or predicted to occur involving cationic or anionic redox remains the subject of much discussion. Although the concept had been examined and discussed earlier, the first conclusive demonstrations of reversible anionic redox effects in transition metal oxide-based Li-ion battery cathode materials and their contribution to battery performance were reported fairly recently by Koga and Sathiya in 2013 [13,20].

A raft of methods, including magnetic susceptibility measurements, have been applied to examine the accommodation of delithiation during charging at transition metal oxide cathodes. In particular, a combined cationic/anionic redox activity in excess Li containing $Li_4Mn_2O_5$ is represented by:

$$Li_4^+Mn_2^{3+}O_5^{2-} \longrightarrow 4Li^+ + 4e^- + Mn_x^{4+}Mn_{2-x}^{5+}O_{5-x}^{2-}O_x^{1-} \qquad (3)$$

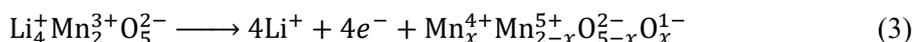

Where $x$ indicates the extent of oxygen participation in redox activity [22]. By density functional theory (DFT) studies reported in this work, it was elucidated that for different levels of lithiation, different redox mechanisms take place, with these mechanisms shedding light on the diminishment of capacity following initial cycling. Initially, during charging, cation oxidation $Mn^{3+}$ to $Mn^{4+}$ takes place, while additional de-lithiation capacity is accommodated by the oxygen oxidation–$O^{2-}$ to $O^{1-}$, and finally, at low levels of Li in $Li_xMnO_5$ ($x<1$), the oxidation of $Mn^{4+}$ to $Mn^{5+}$ takes place, with associated change from octahedral to tetrahedral coordination of this species.

### 3.1 Contribution to Capacity vs. Detrimental $O_2$ evolution

The redox activity of transition metal cations occurring in most TMO-based battery materials used today arises when the Fermi level is located in the MO* band, the band derived from M–O antibonding electronic levels. In these materials, the involvement or contribution of the oxygen anion to the redox band is considerably weaker than the cationic contribution but nevertheless non-negligible. The weak involvement of anions in redox occurs through lowering the electrochemical potential. The activation of anionic redox is strongly correlated to the energies of the aforementioned TMO* antibonding orbitals and the O $2p$ orbitals. Non-bonding O $2p$ states, also known as orphaned O $2p$ states, are widely considered to be a prerequisite for the occurrence of oxygen redox effects in transition metal oxides [23]. In sodium-ion battery systems, the Na–O–Na bonds in excess alkali materials favor the formation of these non-bonding orbitals. Commonly, the activation mediated by these orphaned oxygen states occurs under deep charge conditions, such as lithium de-intercalation under high voltages [24,25]. In experimental investigations of both sodium- and lithium-ion batteries, this manifests by the high first-cycle voltage charge/discharge plateaus seen for materials where this occurs. These plateaus are observed at lower voltages in materials where oxygen redox occurs more readily and are associated with voltage fade and lower corresponding long-term capacities. The observed degradation in capacity may also be associated with structural changes, often a transformation to a spinel phase [26,27].

While it is evidently and theoretically shown to produce greater capacities through greater de-intercalation of lithium or sodium, the oxidation of oxygen in transition metal oxides lacking sufficient covalency between metals and oxygen atoms may result in the diffusion of transition metals and the release of oxygen gas near particle surfaces [19,25]. Following the first charge/discharge cycle, these irreversible reactions decay the cathode material's structure and capacity. This is a known problem in oxygen-redox-driven transition metal oxide-based cathodes in lithium and sodium ion battery systems [28]. Oxygen evolution, which results from some of these types of reactions, although mainly limited to surface areas, poses a significant ignition hazard, resulting in unacceptable flammability risks for cathode materials in which these irreversible oxygen redox-mediated decompositions occur. Along with a large first-cycle capacity loss, this is one of the factors limiting the real-world utilization of Li-rich layered manganese oxide cathode materials, in which lithium extraction and insertion are enabled by oxygen redox phenomena[28]. To harness anionic redox to extract greater capacity from a cathode while minimizing oxygen evolution and irreversible performance-degrading transformations, surface modification can be carried out to inhibit oxygen loss [18,29]. Surface fluorination has emerged in recent years as an effective method for suppressing oxygen evolution and the associated capacity loss [28,30,31]. Various DFT studies have investigated the mechanisms through which surface fluorination methods increase the energetic barrier to oxygen vacancy formation. In conjunction with X-ray photoelectron spectroscopy, nuclear magnetic resonance., and electrochemical measurements, these theoretical studies have shown that surface localized metal-fluoride bonds strengthen TM–O bonds in this region and reduce Jahn-Teller distortion, thus making oxygen sites stable throughout oxygen redox cycles [28,32-34]. Fluorination of grain surfaces to facilitate beneficial anionic redox activity requires determining the optimal level of modification, as with such methods, there is a tradeoff between initial capacity diminishment and long-term capacity retention [34].

### 3.2 Computational Perspective and Future Direction

The ubiquity of anionic redox activity in the broad range of transition metal oxides studied for use as battery cathodes is the subject of considerable scrutiny [35,36]. The various modalities through which oxygen may contribute to the redox activity in Li-excess or Na-excess layered oxides can be interpreted and studied using various continually evolving computational approaches [37-39]. The range of possible compositions and structures in layered alkali transition metal oxides means that oxygen redox activity needs to be studied in diverse systems, experimentally and computationally. The energy levels of the possible transition metal and oxygen states will determine which of these species undergo oxidation when $Na^+/Li^+$ are removed from the cathode during charging of the battery and to what extent. These energy levels are strongly influenced by bonding and the local environment in a lattice structure that is generally non-stoichiometric.

The application of computational methods to understand anionic redox activity in general, particularly in layered transition metal oxides, offers a cost-effective approach to potentially extract higher performance levels from legacy materials, such as LMNOs, or to screen new candidate materials [40]. Such computational studies can be conducted to identify the cationic speciation most conducive to enhancing anionic redox behavior, allowing the maximization of capacities beyond what would be expected solely based on cationic redox pathways. This aspect of lithiation and de-lithiation is studied frequently through DFT- and molecular dynamics-based tools [41].

Technically, examining anionic redox contribution in cathode materials using density functional calculations is straightforward, albeit with a caveat. As seen in Fig. 2d, O 2*p* states, when coordinated with more than three Li (or equivalently Na), are lifted closer to the Fermi level and thus become labile for redox reaction.

Accurate density functional calculations that produce a correct density of states profile for O $2p$ and transition metal $3d$ or $4d$ states can predict the extent of O's participation in the redox reaction. Such participation is calculated by integrating the O's $2p$ states over a narrow window below the Fermi level, typically 2.0 ~ 2.5 eV [15,42], in the lithiated (or sodiated) material to calculate the number of labile oxygen electrons for the redox reaction. Note that states closer to the Fermi level are the first to vacate during the reduction accompanied by the removal of the alkali ions.

For any theoretical method to accurately assess the participation of O ions in the redox reaction, the correct location of the TM and O states with respect to one another and with respect to the Fermi level is paramount. However, many common implementations of exchange-correlation functionals in density functional theory fail to reach the desired accuracy. For instance, the most commonly used family of functionals based on general gradient approximation (GGA) [43], including Perdew-Burke-Ernzerhof (PBE) functional [44,45], the revised Perdew-Burke-Ernzerhof (revPBE) functional [46], and the Becke-Johnson (BJ) functional [47] overly delocalise the $3d$ states, inaccurately prediction wider bands and narrower band gaps. GGA's inaccuracy stems from the assumption that the electrons in the system can be treated as independent, non-interacting entities. However, this assumption is not valid in strongly correlated systems, which include $3d$ and $4d$ transition metal oxides. As such, GGA alone is not suitable for investigating ionic redox.

To remedy GGA's shortcomings, one ought to implement higher-level corrections to the GGA scheme. The Hartree-Fock (HF) hybrid functional [48] is one popular option. HF hybrid functional combines the exact exchange from Hartree-Fock (HF) method with the more general functional, typically GGA. The percentage of the HF exchange energy mixed with GGA is to be determined by a mixing parameter $\alpha$. The Hartree-Fock method is a type of quantum mechanical calculation that explicitly models the interactions between electrons in a many-electron system, thus surpassing the GGA accuracy, albeit at a higher (~10 folds for a typical oxide unitcell) computational cost. In particular, the portion calculated using Hartree-Fock accounts for the electron-electron interactions in the system more sophisticatedly than is possible with pure DFT, while the portion calculated using DFT accounts for the interactions between the electrons and the ions in the system.

GGA+$U$ is another widely used computational method in density functional theory that has been shown to provide a more accurate description of the electronic properties in strongly correlated materials. The GGA+$U$ approach attempts to correct the standard GGA exchange-correlation functional for self-interaction error by introducing an *ad hoc* onsite Coulomb interaction $U$ in narrow spatially localized bands, i.e., $3d$ or $4d$ bands, based on the Hubbard model approach for treating strongly correlated electrons [49]. Although the band placement in the GGA+$U$ method is superior to that of the GGA method in many applications [50], it is still argued that in $3d$ transition metal oxides, the level of the correction is still not adequate for accurate O redox prediction [15,51]. However, since the extent of electronic localization in $4d$ oxides is not as dominant as in $3d$ oxides, by virtue of being farther from the nucleus, the GGA+$U$ method is effective [42,52].

In both HF hybrid functional and GGA+$U$ methods, empirical parameters $\alpha$ and $U$ representing the strength of the electron-electron interactions within a localized $3d$ or $4d$ orbital must be realistically determined for accurate redox calculations. In HF hybrid functional, $\alpha$ can be determined by benchmarking the density of states calculated with the accurate but computationally intensive GW method that accounts for the self-energy of many-body electronic systems [53]. Once a suitable $\alpha$ is determined for a sample system, that $\alpha$ value can be relatively safely applied to materials of similar composition that manifest during the charge/discharge cycle [15]. For the GGA+$U$ method, the value of $U$ can be determined empirically by comparing the results of calculations with experimental data or by fitting the value to the system under study [17,54]. There are also various methods for calculating $U$ based on first principles, such as the linear response approach [55] or the constrained random phase approximation [56]. The choice of the appropriate $\alpha$ and $U$ ultimately determines the desired accuracy level.

## 4 Oxygen Redox in 3d and 4d Ilmenite-type Na$_x$TMO$_3$

Here, we review the O redox mechanism in ilmenite-type oxides with the general formula NaTMO$_3$. Interestingly, O redox in this class of materials is not caused by Na over-coordinating oxygen but rather by O being under-coordinated. First, let's examine the ilmenite structure, named after the mineral ilmenite FeTiO$_3$. Ilmenite's crystal structure, shown in Fig. 3a, is classified as rhombohedral, specifically in the space group $R\bar{3}$ (No. 148) according to the International Tables for Crystallography. Ilmenite has a conventional hexagonal cell with three formula units per cell, in which metal ions are arranged in a close-packed oxygen lattice, with oxygen atoms forming a distorted octahedral coordination around each cation. Here, similar to Li or Na excess structures, O ions are coordinated by two TM ions only, along with two Na ions. Although ilmenite-type structures are considered Na deficient, TM under-coordination creates unhybridized non-bonding O 2$p$ states situated just below the Fermi level, which are labile for anionic redox.

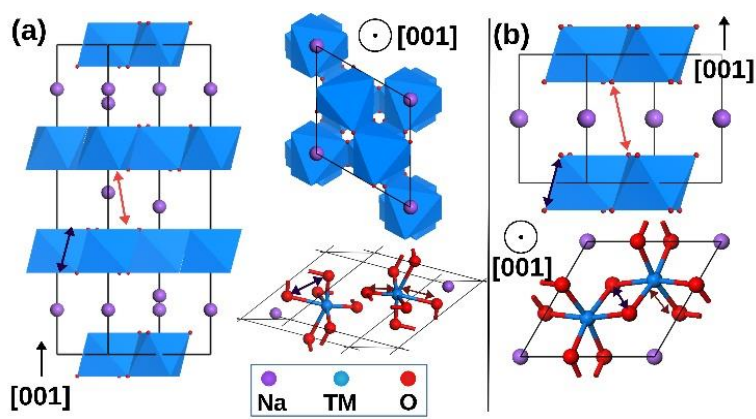

**Fig. 3. (a) The lattice structure of ilmenite-type NaTMO$_3$ compounds. The polyhedral model represents the conventional cell, while the ball and stick model represents the primitive cell. NaTMO$_3$ de-sodiates to Na$_{0.5}$TMO$_3$, shown in (b). Adapted with permission from Ref. [51]. Copyright 2020 American Physical**

The motivation for studying ilmenite-type Na TM oxides was the observation of a relatively high capacity of 180 mAhg$^{-1}$ in Na$_{2-x}$RuO$_3$ (0 < $x$ < 1.5) corresponding to a redox of ~1.4 electrons per formula unit [57]. Here, the additional ~0.4 electron capacity was attributed to a reversible oxygen redox mechanism after the compound was de-sodiated to ilmenite-type Na$_1$RuO$_3$. Furthermore, after the oxygen redox mechanism was initiated, Na$_{1-x}$RuO$_3$ exhibited a voltage spike from 2.2 V to 3.4 V. The observation of the higher voltage hinted at the fact that O redox might have generated a higher operating potential than the TM redox alone. Accordingly, a comprehensive survey spanning ilmenite-type structures with 3$d$ (TM = V, Cr, Mn, Fe, Co, and Ni) and 4$d$ (TM = Nb, Mo, Tc, Ru, Rh, Pd, and Ag) NaTMO$_3$ was conducted with density functional calculations [42,51]. To calculate the operating voltage, NaTMO$_3$ was de-sodiated to Na$_{0.5}$TMO$_3$ with a hexagonal ($P\bar{3}1m$) structure, shown in Fig. 3b. The voltage was calculated according to the following equation:

$$V = \frac{[E^t(\mathrm{Na}_1\mathrm{TMO}_3) - E^t(\mathrm{Na}_{0.5}\mathrm{TMO}_3) - E^t(\mathrm{Na})]}{-e}. \quad (4)$$

Here, $E^t(\mathrm{Na}_x\mathrm{TMO}_3)$ is the total DFT energy for Na$_x$TMO$_3$, $E^t(\mathrm{Na})$ is the total energy of the solid metallic Na per atom, and $e$ is the electron's charge.

Computationally, For the 3$d$ containing NaTMO$_3$ compounds, HF hybrid functional, with a mixing parameter of $\alpha$ = 0.2, was used to calculate the electronic properties, while for the 4$d$ containing compounds, the GGA+$U$ approach with $U_{\mathrm{eff}}$ = 2.0 eV was used. $U_{\mathrm{eff}}$ and $\alpha$ values were kept constant across the 3$d$, and 4$d$ compound sets to make property trends detectable. Fig. 4 shows the calculated density of states across the investigated compounds. A quick visual inspection reveals that in these ilmenite-type compounds, perhaps with an exception for NaMoO$_3$, O 2$p$ states dominate the energy bracket below the Fermi level. This orbital arrangement is in stark contrast to a typical NaTMO$_2$ com pound in which O 2$p$ states tend to gravitate towards the bottom of the valence band [16,58,59].

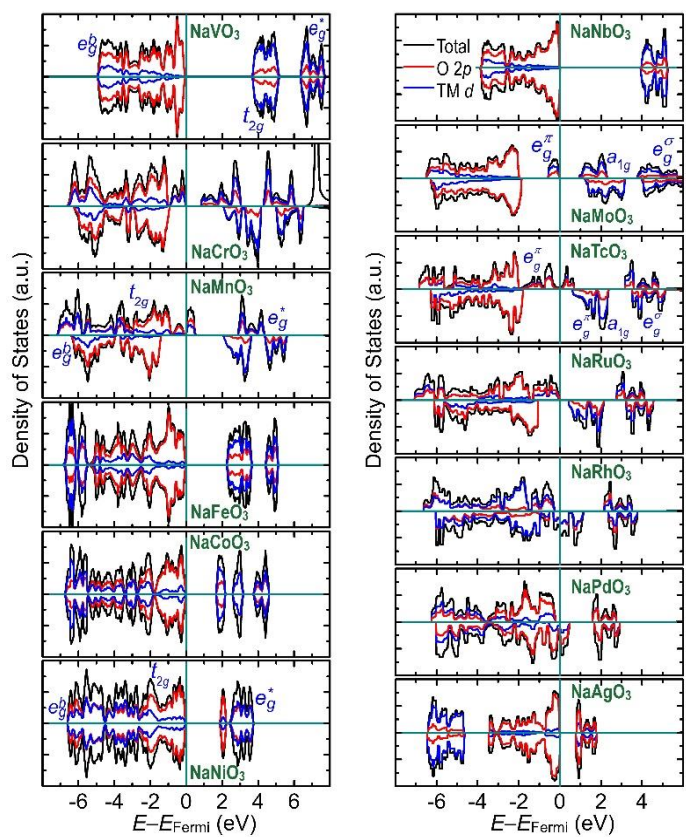

**Fig. 4.** The density of states for NaTMO$_3$ compounds. The left column contains the 3$d$ TM compounds, while the right column contains the 4$d$ TM compounds. Black, red, and blue lines represent the total, O 2$p$, and TM $d$ states, respectively. As these DOS diagrams are spin-polarized, positive DOS values denote spin-up states, while negative values denote spin-down states. Adapted with permission from Refs. [42,51]. Copyrights 2020 American Physical Society and 2018 Royal Society of Chemistry.

Table 1 and Table 2 show the labile O 2$p$ population and the calculated operating potential of the NaTMO$_3$ cathodes. Noticeably, the operating potential of the 3$d$ TM-containing compounds is rather large. Especially, NaVO$_3$ and NaFeO$_3$ were predicted to operate at voltages higher than 5 V. Among the 4$d$ TM-containing compounds, NaNbO$_3$ and NaAgO$_3$ were similarly predicted to operate at potentials higher than 4 V. In the case of NaRuO$_3$, the calculated potential of 3.27 V agrees rather well with the observed value of ~3.4 V [57], confirming the reliability of the calculations.

Furthermore, all compounds, with varying degrees, exhibit the availability of their O 2$p$ states for redox. NaCrO$_3$, NaMnO$_3$, NaNbO$_3$, NaRuO$_3$, NaRhO$_3$, and NaAgO$_3$ all have more than one O 2$p$ electron within the redox energy window, and NaCoO$_3$ and NaNiO$_3$ have more than half O 2$p$ electron. Apparently, the 4$d$ TM-containing compounds, in general, have a higher labile oxygen population. This trend may stem from the higher covalency in 4$d$ TM bonds with oxygen that tends to substantially increase the 4$d$–2$p$ hybridization over the entire valence band, and therefore, increasing the O 2$p$ population near the Fermi level at the expense of the O 2$p$ population at the bottom of the valence band [60].

Table 1. The O 2$p$ electronic population within the energy interval of $-2$ eV $< E < E_{Fermi}$, and the calculated operating potential for 3$d$ containing ilmenite-type NaTMO$_3$ compounds.

| Compound | NaVO$_3$ | NaCrO$_3$ | NaMnO$_3$ | NaFeO$_3$ | NaCoO$_3$ | NaNiO$_3$ |
|---|---|---|---|---|---|---|
| O 2$p$ populations ($e$) | 0.17 | 1.43 | 1.23 | 0.34 | 0.59 | 0.78 |
| Potential (V) | 5.91 | 4.24 | 4.53 | 5.18 | 3.49 | 3.98 |

Table 2. The O 2$p$ electronic population within the energy interval of $-2$ eV $< E < E_{Fermi}$, and the calculated operating potential for 4$d$ containing ilmenite-type NaTMO$_3$ compounds.

| Compound | NaNbO$_3$ | NaMoO$_3$ | NaTcO$_3$ | NaRuO$_3$ | NaRhO$_3$ | NaPdO$_3$ | NaAgO$_3$ |
|---|---|---|---|---|---|---|---|
| O 2$p$ populations ($e$) | 2.58 | 0.16 | 0.38 | 1.40 | 1.50 | 2.10 | 2.51 |
| Potential (V) | 4.95 | 2.62 | 2.90 | 3.72 | 3.63 | 4.08 | 4.66 |

## 5 Summary

The review of the anionic, or oxygen, redox in transition metal oxide revealed the wide occurrence of this redox mechanism in the common cathode materials for rechargeable Li and Na ion batteries. However, the extent of this redox mechanism varies based on the crystal structure and the type of transition metal. Initially, oxygen redox was discovered to significantly contribute toward the overall battery capacity in Li-excess disordered spinel-type cathodes. Here, the high O redox contribution is attributed to orphaned oxygen states arising from those oxygen ions with less than three coordinating transition metal elements. Later, significant O redox was observed in ilmenite-type oxides in which oxygen atoms were under-coordinated by both alkali and transition metal ions. In this case, vacancies around the O ions create orphaned O states. In both spinel- and ilmenite-type oxides, the orphaned O 2$p$ states are closer to the Fermi level than the fully hybridized O states and therefore are more labile for the redox reaction. Furthermore, the O redox reaction was found to be more dominant in oxides containing 4$d$ (fifth row) transition metal oxides. Although anionic redox increases the cathode capacity, the reduced O ions have a strong tendency towards bonding to each other and leaving the lattice as O$_2$ gas. In addition to being a fire hazard, O$_2$ release during battery operation causes rapid and permanent capacity fade. Consequently, careful mitigation strategies against O$_2$ release, such as fluorination, must be implemented should the extra capacity from O redox be utilized. Computationally, O redox can be, in principle, predicted by analyzing the density of states obtained from density functional calculations. However, not all types of calculations are equally accurate. Especially the type of functional chosen must produce a correct band structure for the cathode material. Given that most cathode materials fall under the classification of strongly correlated materials, standard general gradient approximation certainly falls short in correctly predicting the O redox. Higher level functionals such as Hartree-Fock hybrid functional and occasionally Hubbard-corrected functionals must be considered for the density functional calculations probing O redox. However, identifying adequate values for phenomenological constants, such as the mixing parameter or the strength of the on-side Hubbard term, remains a challenge when using these higher-level functionals.

## References

[1] D.-H. Seo, A. Urban, and G. Ceder, Phys. Rev. B **92** (2015).
[2] A. Massaro, A. B. Munoz-Garcia, P. P. Prosini, C. Gerbaldi, and M. Pavone, ACS Energy Lett. **6**, 2470 (2021).
[3] A. Michail, B. Silván, and N. Tapia-Ruiz, Curr. Opin. Electrochem. **31**, 100817 (2022).
[4] M. H. N. Assadi, C. J. Kirkham, I. Hamada, and D. A. Hanaor, Chem. Phys. Lett., 139694 (2022).

[5]     M. Okubo and A. Yamada, ACS Appl. Mater. Interfaces **9** (2017).
[6]     B. Ammundsen, J. Desilvestro, T. Groutso, D. Hassell, J. Metson, E. Regan, R. Steiner, and P. Pickering, J. Electrochem. Soc. **147**, 4078 (2000).
[7]     B. L. Ellis and L. F. Nazar, Curr. Opin. Solid State Mater. Sci. **16**, 168 (2012).
[8]     C. Didier, M. Guignard, C. Denage, O. Szajwaj, S. Ito, I. Saadoune, J. Darriet, and C. Delmas, Electrochem. Solid-State Lett. **14**, A75 (2011).
[9]     L. de Biasi, A. Schiele, M. Roca-Ayats, G. Garcia, T. Brezesinski, P. Hartmann, and J. Janek, ChemSusChem **12**, 2240 (2019).
[10]    J. Young and M. Smeu, Adv. Theory Simul. **4** (2021).
[11]    S. Sharifi-Asl, J. Lu, K. Amine, and R. Shahbazian-Yassar, Adv. Energy Mater. **9** (2019).
[12]    U. Maitra, R. A. House, J. W. Somerville, N. Tapia-Ruiz, J. G. Lozano, N. Guerrini, R. Hao, K. Luo, L. Jin, and M. A. Pérez-Osorio, Nat. Chem. **10**, 288 (2018).
[13]    H. Koga, L. Croguennec, M. Ménétrier, K. Douhil, S. Belin, L. Bourgeois, E. Suard, F. Weill, and C. Delmas, J. Electrochem. Soc. **160**, A786 (2013).
[14]    M. Ben Yahia, J. Vergnet, M. Saubanère, and M.-L. Doublet, Nat. Mater. **18**, 496 (2019).
[15]    D.-H. Seo, J. Lee, A. Urban, R. Malik, S. Kang, and G. Ceder, Nat. Chem. **8** (2016).
[16]    M. Aydinol, A. Kohan, G. Ceder, K. Cho, and J. Joannopoulos, Phys. Rev. B **56**, 1354 (1997).
[17]    M. H. N. Assadi and Y. Shigeta, J. Power Sources **388**, 1 (2018).
[18]    E. Hu, X. Yu, R. Lin, X. Bi, J. Lu, S. Bak, K.-W. Nam, H. L. Xin, C. Jaye, and D. A. Fischer, Nat. Energy **3**, 690 (2018).
[19]    E. Song, Y. Hu, R. Ma, Y. Li, X. Zhao, J. Wang, and J. Liu, J. Mater. Chem. A **9** (2021).
[20]    M. Sathiya, G. Rousse, K. Ramesha, C. Laisa, H. Vezin, M. T. Sougrati, M.-L. Doublet, D. Foix, D. Gonbeau, and W. Walker, Nat. Mater. **12** (2013).
[21]    V. Murugan, R. S. Arul Saravanan, K. Thangaian, T. Partheeban, V. Aravindan, M. Srinivasan, M. Sasidharan, and K. K. Bharathi, ACS Appl. Energy Mater. **4** (2021).
[22]    Z. Yao, S. Kim, J. He, V. I. Hegde, and C. Wolverton, Sci. Adv. **4** (2018).
[23]    Q. Wu, T. Zhang, J. Geng, S. Gao, H. Ma, and F. Li, Energy Fuels **36** (2022).
[24]    Q. Liu, Z. Hu, W. Li, C. Zou, H. Jin, S. Wang, S. Chou, and S.-X. Dou, Energy Environ. Sci. **14**, 158 (2021).
[25]    Y. Yu, P. Karayaylali, D. Sokaras, L. Giordano, R. Kou, C.-J. Sun, F. Maglia, R. Jung, F. S. Gittleson, and Y. Shao-Horn, Energy Environ. Sci. **14** (2021).
[26]    Y. Lee, J. Shin, H. Kang, D. Lee, T.-H. Kim, Y.-K. Kwon, and E. Cho, Adv. Sci. **8**, 2003013 (2021).
[27]    W. Yang, Nat. Energy **3** (2018).
[28]    T. Li, X. Xia, J. Liu, Z. Liu, S. Hu, L. Zhang, Y. Zheng, Z. Wang, H. Chen, and M. Peng, Energy Storage Mater. **49**, 555 (2022).
[29]    S. Sharifi-Asl, V. Yurkiv, A. Gutierrez, M. Cheng, M. Balasubramanian, F. Mashayek, J. Croy, and R. Shahbazian-Yassar, Nano Lett. **20** (2019).
[30]    K. Zhou, S. Zheng, F. Ren, J. Wu, H. Liu, M. Luo, X. Liu, Y. Xiang, C. Zhang, and W. Yang, Energy Storage Mater. **32** (2020).
[31]    L. Li, J. Ahn, Y. Yue, W. Tong, G. Chen, and C. Wang, Adv. Mater. **34**, 2106256 (2022).
[32]    L. Zhu, Y. Liu, W. Wu, X. Wu, W. Tang, and Y. Wu, J. Mater. Chem. A **3**, 15156 (2015).
[33]    R. J. Clément, D. Kitchaev, J. Lee, and G. Ceder, Chem. Mater. **30**, 6945 (2018).
[34]    Z. Lun, B. Ouyang, D. A. Kitchaev, R. J. Clément, J. K. Papp, M. Balasubramanian, Y. Tian, T. Lei, T. Shi, and B. D. McCloskey, Adv. Energy Mater. **9**, 1802959 (2019).
[35]    W.-J. Kwak, H. Kim, H.-G. Jung, D. Aurbach, and Y.-K. Sun, J. Electrochem. Soc. **165**, A2274 (2018).
[36]    W. Zheng, G. Liang, S. Zhang, K. Davey, and Z. Guo, Nano Res., 10.1007/s12274-022-5003-1 (2022).
[37]    A. Van der Ven, Z. Deng, S. Banerjee, and S. P. Ong, Chem. Rev. **120**, 6977 (2020).
[38]    X. Chen, X. Liu, X. Shen, and Q. Zhang, Angew. Chem. **133**, 24558 (2021).
[39]    S. Ding, X. Yu, Z.-F. Ma, and X. Yuan, J. Mater. Chem. A **9**, 8160 (2021).
[40]    M. Zhang, D. A. Kitchaev, Z. Lebens-Higgins, J. Vinckeviciute, M. Zuba, P. J. Reeves, C. P. Grey, M. S. Whittingham, L. F. Piper, and A. Van der Ven, Nat. Rev. Mater. **7**, 522 (2022).
[41]    A. S. Tygesen, J. H. Chang, T. Vegge, and J. M. García-Lastra, NPJ Comput. Mater. **6**, 1 (2020).
[42]    M. Assadi, M. Okubo, A. Yamada, and Y. Tateyama, J. Mater. Chem. A **6**, 3747 (2018).
[43]    A. D. Becke, Phys. Rev. A **38**, 3098 (1988).
[44]    J. P. Perdew, K. Burke, and M. Ernzerhof, Phys. Rev. Lett. **77** (1996).
[45]    J. P. Perdew, K. Burke, and M. Ernzerhof, Phys. Rev. Lett. **78** (1997).
[46]    B. Hammer, L. B. Hansen, and J. K. Nørskov, Phys. Rev. B **59**, 7413 (1999).
[47]    A. D. Becke and E. R. Johnson, J. Chem. Phys. **124**, 221101 (2006).
[48]    A. D. Becke, J. Chem. Phys. **98**, 5648 (1993).

[49]  V. I. Anisimov, J. Zaanen, and O. K. Andersen, Phys. Rev. B **44**, 943 (1991).
[50]  A. Peles, J. Mater. Sci. **47** (2012).
[51]  M. Assadi, M. Okubo, A. Yamada, and Y. Tateyama, Phys. Rev. Mater. **4**, 015401 (2020).
[52]  M. Assadi, M. Okubo, A. Yamada, and Y. Tateyama, J. Electrochem. Soc. **166**, A5343 (2019).
[53]  W. G. Aulbur, L. Jönsson, and J. W. Wilkins, in *Solid State Phys.*, edited by H. Ehrenreich, and F. Spaepen (Academic Press, 2000), p. 1–218.
[54]  M. S. Islam and C. A. J. Fisher, Chem. Soc. Rev. **43**, 185 (2014).
[55]  M. Cococcioni and S. de Gironcoli, Phys. Rev. B **71**, 035105 (2005).
[56]  L. Vaugier, H. Jiang, and S. Biermann, Phys. Rev. B **86** (2012).
[57]  B. Mortemard de Boisse, G. Liu, J. Ma, S.-i. Nishimura, S.-C. Chung, H. Kiuchi, Y. Harada, J. Kikkawa, Y. Kobayashi, M. Okubo, and A. Yamada, Nat. Commun. **7**, 11397 (2016).
[58]  H. Li, Z. Wang, L. Chen, and X. Huang, Adv. Mater. **21**, 4593 (2009).
[59]  M. H. N. Assadi and Y. Shigeta, RSC Adv. **8**, 13842 (2018).
[60]  M. H. N. Assadi, M. Fronzi, M. Ford, and Y. Shigeta, J. Mater. Chem. A **6**, 24120 (2018).